\documentclass[a4paper]{article}

\usepackage{INTERSPEECH2022}
\usepackage{xcolor}
\usepackage{booktabs}
\usepackage{url}

\usepackage{multirow}

\newcommand{\todo}[1]{}
\renewcommand{\todo}[1]{{\color{red} TODO: {#1}}}

\title{BUT System for the MLC-SLM Challenge}
\name{Alexander Polok$^1$, Jiangyu Han$^1$, Dominik Klement$^1$, Samuele Cornell$^2$, Jan Černocký$^1$, Lukáš Burget$^1$}

\address{
  $^1$Speech@FIT, Brno University of Technology, Czechia\\
  $^2$Language Technologies Institute, Carnegie Mellon University, USA}
\email{ipoloka@fit.vut.cz}

\makeatletter
\def\bstctlcite{\@ifnextchar[{\@bstctlcite}{\@bstctlcite[@auxout]}}
\def\@bstctlcite[#1]#2{\@bsphack
  \@for\@citeb:=#2\do{%
    \edef\@citeb{\expandafter\@firstofone\@citeb}%
    \if@filesw\immediate\write\csname #1\endcsname{\string\citation{\@citeb}}\fi}%
  \@esphack}
\makeatother 

\begin{document}
\bstctlcite{IEEEexample:BSTcontrol}

\maketitle
\begin{abstract}
We present a two-speaker automatic speech recognition (ASR) system that combines DiCoW—a diarization-conditioned variant of Whisper—with DiariZen, a diarization pipeline built on top of Pyannote. We first evaluate both systems in out-of-domain (OOD) multilingual scenarios without any fine-tuning.
In this scenario, DiariZen consistently outperforms the baseline Pyannote diarization model, demonstrating strong generalization. Despite being fine-tuned on English-only data for target-speaker ASR, DiCoW retains solid multilingual performance, indicating that encoder modifications preserve Whisper’s multilingual capabilities.
We then fine-tune both DiCoW and DiariZen on the MLC-SLM challenge data. The fine-tuned DiariZen continues to outperform the fine-tuned Pyannote baseline, while DiCoW sees further gains from domain adaptation.
Our final system achieves a micro-average tcpWER/CER of 16.75\,\% and ranks second in Task 2 of the MLC-SLM challenge.
Lastly, we identify several labeling inconsistencies in the training data—such as missing speech segments and incorrect silence annotations—which can hinder diarization fine-tuning. We propose simple mitigation strategies to address these issues and improve system robustness.
\end{abstract}


\section{Introduction}
Recent advances in deep learning, particularly the rise of large self-supervised models~~\cite{chen2022wavlm, hsu2021hubert}, large language models (LLMs)~\cite{achiam2023gpt, touvron2023llama}, and Whisper-style supervised architectures~\cite{radford2023robust, peng2023owsm}, have led to substantial progress in automatic speech recognition (ASR), even under challenging acoustic conditions.
These models achieve remarkable accuracy by leveraging massive training data~\cite{li2023yodas, chen21o_interspeech} and scaling up model parameters~\cite{chen2025owlsscalinglawsmultilingual}.
However, such advancements have primarily benefited single-speaker ASR, while most real-world scenarios involve conversational speech between multiple speakers~\cite{cornell2023chime, cornell2024chime, vinnikov24_interspeech}.
Several approaches have been proposed for speaker-attributed ASR in multi-speaker settings~\cite{niu24_chime, polok24_butjhu, huang24b_chime}.
Some systems operate modularly, combining diarization, source separation, speaker clustering, and ASR as separate components.
Others follow end-to-end strategies incorporating speaker tokens~\cite{kanda20b_interspeech, cornell2024one} or use multiple decoder heads to generate separate transcripts~\cite{yu2017permutation}.

As an alternative, Target-speaker ASR offers a middle ground by conditioning ASR models directly on speaker identity using embeddings or enrollment audio~\cite{Kanda2019_spkloss, zhang_23_conformer,Zili23_adapting,meng24c_interspeech}. While effective in controlled settings, these methods often depend on speaker-specific representations, which can be difficult to generalize, especially when training data is limited or speaker variability is low.

To address the limitations of traditional speaker-attributed ASR systems, our previously proposed method—Diarization-Conditioned Whisper (DiCoW)~\cite{polok24_butjhu,polok2024targetspeakerasrwhisper,polok2024dicowdiarizationconditionedwhispertarget}—conditions the model directly on frame-level diarization masks, bypassing the need for explicit speaker identity modeling.

For speaker diarization, self-supervised learning (SSL) models like WavLM~\cite{chen2022wavlm} have shown great potential, as demonstrated by several recent studies~\cite{ plaquet2024mambabasedsegmentationmodelspeaker, han2024leveraging}. Among these, our recent work DiariZen~\cite{han2024leveraging} built upon  Pyannote~\cite{bredin2023Pyannote, plaquet23_interspeech}, achieves competitive performance across diverse benchmarks by combining WavLM with the Conformer~\cite{gulati2020conformer}.

In this work, we build on DiCoW by combining it with DiariZen to form a complete two-speaker ASR pipeline. We begin by evaluating the zero-shot performance of both models on multilingual data—without any domain adaptation—to test their robustness in real-world scenarios. We then fine-tune them on target-domain data to analyze the benefits of adaptation.
Beyond improved performance, we uncover two key insights. First, even though DiCoW was fine-tuned exclusively on English data, it retains Whisper’s multilingual capabilities. This finding aligns with our prior observations~\cite[Table 10]{polok2024dicowdiarizationconditionedwhispertarget}, where the model showed only minimal degradation on standard single-speaker tasks, suggesting that encoder conditioning via diarization masks does not compromise Whisper’s~\cite{radford2023robust} core strengths. Second, during fine-tuning, we observed labeling inconsistencies in the training data—such as skipped speech segments and prolonged silences annotated as speech—which may hinder the effective training of diarization systems. As a result, the diarization model may learn to reproduce these patterns on the target data, negatively impacting downstream ASR performance, even in cases where diarization error rate (DER) on the development set suggests substantial improvements.

To support continued research in this direction, we release both the DiCoW\footnote{\url{https://huggingface.co/BUT-FIT/DiCoW_v3_MLC}} and DiariZen\footnote{\url{https://huggingface.co/BUT-FIT/diarizen-wavlm-large-s80-mlc}} models to the community.

\section{Method}
This section provides an overview of our system, which is designed similarly to the official baseline\footnote{\url{https://github.com/mubingshen/MLC-SLM-Baseline}} but with two key differences: the Pyannote diarization module is replaced by DiariZen, and unlike the baseline system—which processes each chunk extracted from diarization independently, either with Whisper or Whisper connected to an LLM—our approach operates on the full recording and integrates DiCoW. Next, we introduce both DiariZen and DiCoW in detail.

\subsection{DiariZen}
DiariZen is a speaker diarization pipeline built on Pyannote. Given a long audio recording, DiariZen first segments the input into shorter chunks and then applies local end-to-end neural diarization (EEND) to each chunk. For each speaker identified within a chunk by the local EEND, a speaker embedding is extracted from their corresponding speech. These embeddings are then clustered to determine correspondence between speakers across chunks and to generate the final diarization results.

\begin{figure}[t]
    \centering
    \includegraphics[width=1\columnwidth]{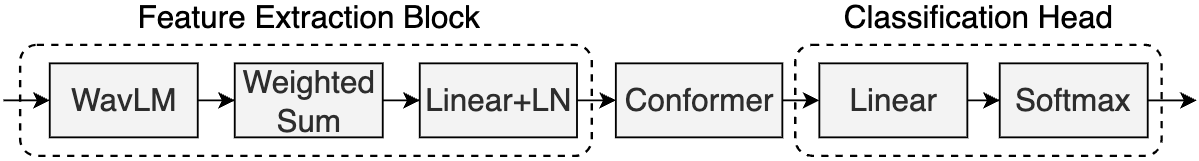}
    \caption{
        Framework of local EEND module for DiariZen. Figure adapted from~\cite{han2024leveraging}.
    }
\label{fig:diarizen}
\end{figure}

As illustrated in Figure \ref{fig:diarizen}, the EEND module in DiariZen combines WavLM and Conformer. The outputs from each WavLM layer are aggregated using a weighted sum to form the input sequence for the Conformer. A classification head is then used to map the Conformer outputs to powerset states~\cite{plaquet23_interspeech}.

\begin{figure}[t]
    \centering
    \includegraphics[width=0.7\columnwidth]{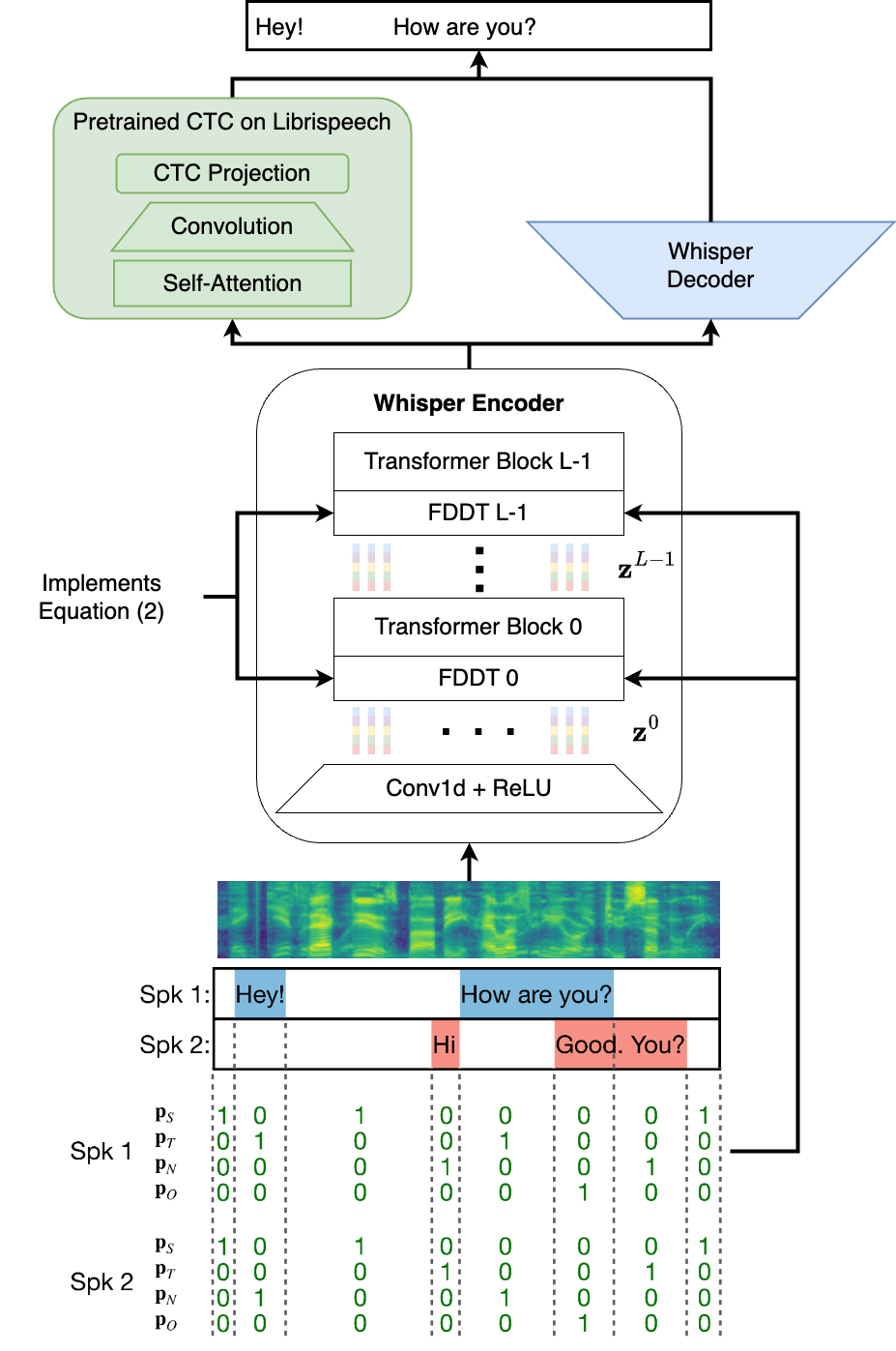}
    \caption{
        Overview of the DiCoW model architecture.
        The model is based on the Whisper architecture, with modifications to incorporate frame-level diarization information through Frame-Level Diarization Dependent Transformations (FDDT).
        Figure adapted from~\cite{polok2024dicowdiarizationconditionedwhispertarget}.
    }
\label{fig:dicow}
\vspace{-10pt}
\end{figure}

\subsection{DiCoW}
DiCoW extends the Whisper architecture for target-speaker ASR (TS-ASR) by conditioning the model on frame-level diarization signals, as illustrated in Figure~\ref{fig:dicow}. Unlike traditional TS-ASR approaches that rely on speaker embeddings or enrollment utterances, DiCoW uses probabilistic speaker activity estimates derived from diarization. Specifically, it computes a Silence-Target-NonTarget-Overlap (STNO) mask that captures the likelihood of four speaking conditions at each time frame: silence, target speaker active, non-target speaker active, and overlap of target speaker with different speaker(s). These probabilities are defined as follows:
\begin{align}\label{eq:stno}
    p_{\mathcal{S}}^t  &= \prod_{s=1}^S (1 - d(s, t)), \quad
    p_{\mathcal{T}}^t  = d(s_k, t)  \prod_{\substack{s=1 \\ s \neq s_k}}^S (1 - d(s, t)) \nonumber \\
    p_{\mathcal{N}}^t  &= \left(1 - p_{\mathcal{S}}^t\right) - d\left(s_k, t\right), \quad
    p_{\mathcal{O}}^t  = d(s_k, t) - p_{\mathcal{T}}^t,
\end{align}
where $S$ is a number of speakers, and $d(s,t)$ is probabilty of speaker $s$ speaking in timestamp $t$. Each Transformer layer in the Whisper encoder layer is then augmented with four affine transformation matrices—one for each STNO class—which are blended based on their probabilities for each frame, forming the Frame-Level Diarization-Dependent Transformations (FDDT):
\begin{align}
\label{eq:FDDT_summary}
\hat{\mathbf{z}}^l_t = &\left( \mathbf{W}_{\mathcal{S}}^l \mathbf{z}^l_t + \mathbf{b}_{\mathcal{S}}^l \right) p^t_{\mathcal{S}} +
\left( \mathbf{W}_{\mathcal{T}}^l \mathbf{z}^l_t + \mathbf{b}_{\mathcal{T}}^l \right) p^t_{\mathcal{T}}  \nonumber \\
 &+ \left( \mathbf{W}_{\mathcal{N}}^l \mathbf{z}^l_t + \mathbf{b}_{\mathcal{N}}^l\right) p^t_{\mathcal{N}} +
\left( \mathbf{W}_{\mathcal{O}}^l \mathbf{z}^l_t + \mathbf{b}_{\mathcal{O}}^l \right) p^t_{\mathcal{O}}.
\end{align}
The FDDT parameters are initialized such that the data flow in the original model is not disrupted~\cite{polok2024targetspeakerasrwhisper}, ensuring that the behavior of the pre-trained Whisper model is preserved at the start. The entire system, including the new FDDT parameters, is then jointly fine-tuned.


\section{Experimental Setup}

\subsection{DiariZen}
Our diarization system builds upon the DiariZen framework\footnote{\url{https://github.com/BUTSpeechFIT/DiariZen}}, following the training approach described in~\cite{han2024leveraging}. We use WavLM Large as the backbone for frame-level classification to improve local modeling. The model is pre-trained on far-field, single-channel audio from a diverse collection of public datasets, including AMI~\cite{Mccowan2005_ami}, AISHELL-4~\cite{fu2021aishell}, AliMeeting~\cite{yu2022m2met}, NOTSOFAR-1~\cite{vinnikov2024notsofar}, MSDWild~\cite{liu2022msdwild}, DIHARD3~\cite{ryant2020third}, RAMC~\cite{yang2022open}, and VoxConverse~\cite{chung2020spot}. 
We then apply structured pruning~\cite{han2025fine} to remove 80\% of WavLM’s parameters. After that, the model is fine-tuned on the official MLC-SLM challenge dataset. We use powerset loss~\cite{plaquet23_interspeech} with two speakers.






We extract speaker embeddings using a ResNet34-LM model trained with the WeSpeaker toolkit~\cite{wang2024advancing} on the VoxCeleb2 dataset~\cite{chung2018voxceleb2}. The embeddings are then clustered using agglomerative hierarchical clustering.

\subsection{DiCoW}
Our training follows the DiCoW~\cite{polok2024dicowdiarizationconditionedwhispertarget} training recipe\footnote{\url{https://github.com/BUTSpeechFIT/TS-ASR-Whisper}} without altering the core hyperparameters or schedule. It uses a three-phase training strategy: CTC preheat, FDDT preheat, and full fine-tuning. For CTC preheating, we utilize LibriSpeech~\cite{librispeech}. During the following two phases, we train Whisper large-v3-turbo on a mix of publicly available multi-speaker datasets—AMI, NOTSOFAR-1, and Libri2Mix~\cite{Cosentino2020LibriMixAO}. Afterwards, we fine-tune the model on the MLC-SLM dataset.

Similar to the original data preparation, we segment the training split into chunks of at most 30 seconds, in accordance with Whisper's input length limitation. To achieve this, we loop through the provided segments and iteratively concatenate consecutive ones until the total duration reaches 30 seconds.


During training, we aim to be more tolerant of differences in annotation styles between the original Whisper training data and the fine-tuning (FT) data. To address this, we apply a challenge text normalization procedure to the reference transcripts by lowercasing all tokens and removing punctuation. For each training sample, we then compute the cross-entropy loss twice—once using the normalized (lowercased and punctuation-removed) tokens, and once using the original cased tokens—and backpropagate the smaller of the two loss values.

We apply early stopping with a patience of 5, selecting the best-performing checkpoint based on tcpWER evaluated every 500 steps using greedy decoding. Training is capped at 50k steps. Final outputs on the test set are generated using beam search with a beam size of 10. Joint CTC decoding~\cite{hori_joint_2017} with a weight of 0.2 is utilized for English utterances. All experiments are conducted on four NVIDIA RTX A6000 GPUs with mixed-precision training and an overall batch size of 96 samples.

\section{Results}
In this section, we evaluate our system’s performance. We begin with diarization error rates (DERs), followed by TS-ASR results using both ground truth and DiariZen-derived segmentations. We then analyze labeling inconsistencies in the training and development data and their impact, and to address these issues, we incorporate an auxiliary VAD model that jointly models speech and silence alongside our diarization system.

\begin{table}[t]
\centering
\caption{Diarization performance of Pyannote and DiariZen-based systems across multiple languages using the MLC-SLM development data. Fine-tuned Pyannote results are from the official baseline.}
\label{tab:exp_diar}
\begin{tabular}{l|cc|ccc}
\toprule
 \multirow{2}{*}{Language}  & \multicolumn{2}{c|}{Pyannote} & \multicolumn{2}{c}{DiariZen} \\
 &  OOD & FT & OOD & FT \\
\midrule
American English              & 33.4 & 20.2 & 24.8 & 15.9  \\
Australian English            & 21.3 & 13.8  & 18.7 & 10.8  \\
British English               & 27.7 & 18.9 & 22.7 & 12.1  \\
Filipino English              & 18.6 & 13.2 & 18.0 & 10.3  \\
Inidian English               & 11.3 & 8.2  & 10.6 & 6.0  \\
French                      & 38.5 & 22.6  & 28.3 & 17.3  \\
German                      & 34.3 & 22.3  & 30.4 & 16.4  \\
Italian                    & 15.2 & 10.6  & 12.4 & 8.9  \\
Japanese                     & 42.0 & 26.5  & 34.7 & 17.8  \\
Korean                     & 42.0 & 23.3  & 24.2 & 16.4  \\
Portuguese                 & 30.2 & 17.6  & 28.3 &  14.8  \\
Russian                   & 16.4 & 11.4  & 12.8 & 10.0  \\
Spanish                  & 23.2 & 12.9  & 21.6 &  10.8  \\
Thai                        & 27.0 & 10.9  & 18.8 & 10.6  \\
Vietnamese                & 27.0 & 14.6  & 20.5 & 12.7  \\
\midrule
\textbf{Overall}   & 27.2 & 16.4 & 21.8 & 12.7  \\
\bottomrule
\end{tabular}
\end{table}

\subsection{Diarization Improvements}
We compare the diarization performance of DiariZen and the baseline Pyannote model under out-of-domain (OOD) and fine-tuned (FT) conditions. Both models are fine-tuned on the MLC-SLM training data, and DER is evaluated without a forgiveness collar. Results are presented in Table~\ref{tab:exp_diar}.
DiariZen consistently outperforms the Pyannote baseline in both OOD and fine-tuned settings, achieving a DER of 12.7\,\% after fine-tuning versus 16.4\,\% for Pyannote.


However, we caution the reader against interpreting this as evidence that diarization fine-tuning itself led to the improvement. These DER gains may be misleading, as they are likely driven by the system learning to better mimic the inconsistent annotation style present in both the training and development data, rather than genuine improvements in diarization quality. We further examine and discuss these issues in Section~\ref{sec:annot}.

\begingroup
\setlength{\tabcolsep}{1.5pt}
\begin{table}[t]
\centering
\caption{tcpWER/CER (\%) on the MLC-SLM development set, broken down by language, comparing the baseline and DiCoW systems using ground-truth (GT) and real diarization before and after fine-tuning (FT). The baseline system uses Whisper large-v3 with chunked inference and a fine-tuned Pyannote diarization model. In contrast, our system uses a fine-tuned DiariZen diarization model. Results marked with an asterisk (*) are reported using tcpCER, following the official evaluation protocol.}
\label{tab:mlc_by_language}
\begin{tabular}{l|ccc|ccc}
\toprule
\multirow{2}{*}{Language}  & \multicolumn{3}{c|}{GT segmentation} & \multicolumn{3}{c}{Real diar} \\
 & Baseline & DiCoW & FT & Baseline & DiCoW & FT  \\
\midrule

American En. & 14.1       & 20.6     & 11.1 & 53.7 & 36.5 & 22.5 \\
Australian En. & 11.7     & 19.4     & 7.4  & 52.6 & 23.6 & 13.0 \\
British En.  & 10.1      & 16.7     & 7.7  & 71.9 & 26.1 & 17.6 \\
Filipino En. & 9.2       & 17.7     & 7.5  & 50.4 & 25.5 & 15.2 \\
Indian En.   & 14.0     & 14.3     & 13.3 & 70.7 & 14.9 & 14.0 \\
French       & 28.1          & 27.7     & 16.1 & 96.0 & 37.8 & 27.5 \\
German       & 20.7          & 21.2     & 23.9 & 86.7 & 30.1 & 27.3 \\
Italian      & 17.9          & 16.2     & 12.3 & 83.3 & 19.8 & 16.4 \\
$\text{Japanese}^{*}$     & 21.6          & 19.2     & 13.7 & 71.3 & 25.8 & 23.3 \\
$\text{Korean}^{*}$       & 13.8          & 12.8     & 8.5  & 59.6 & 24.5 & 22.8 \\
Portuguese   & 21.2          & 24.5     & 19.5 & 118.8 & 33.1 & 29.7 \\
Russian      & 17.7          & 17.6     & 11.6 & 69.2 & 22.5 & 16.7 \\
Spanish      & 12.3          & 11.6     & 8.7  & 75.6 & 18.2 & 16.3 \\
$\text{Thai}^{*}$         & 14.5          & 31.9     & 14.2 & 83.6 & 34.4 & 20.1 \\
Vietnamese   & 27.2          & 30.0     & 15.3 & 82.8 & 33.8 & 24.7 \\

\midrule
\textbf{Overall}   & 16.8    & 22.0     & 12.9 & 76.1 & 28.4 & 20.8  \\
\bottomrule
\end{tabular}
\end{table}
\endgroup

\subsection{Full System Comparison: Diarization and ASR}
In Table~\ref{tab:mlc_by_language}, we compare the overall performance of our system against the Pyannote \& Whisper large-v3 baseline on the MLC-SLM development set. For ground-truth (GT) segmentation (left side of the table), the baseline uses Whisper large-v3, while our system employs Whisper large-v3-turbo adapted for TS-ASR. It is important to note that training TS-ASR exclusively on English data may have limited performance on other languages. Additionally, the turbo model is reported to perform slightly worse than the v3 model on certain languages, such as Thai\footnote{\url{https://github.com/openai/whisper/discussions/2363}}. However, our system’s use of long-form inference may offer compensatory benefits compared to the baseline. After fine-tuning, our model outperforms the large-v3 baseline on GT segmentation. We note that results on German might be affected by labeling inconsistencies in the dataset.

On the right side of Table~\ref{tab:mlc_by_language}, full system results are presented for real diarization segmentations using both the baseline Pyannote \& Whisper large-v3 system and our out-of-domain (OOD) and fine-tuned (FT) DiCoW system conditioned by FT DiariZen segmentation. DiCoW paired with DiariZen without domain adaptation significantly outperforms the baseline Pyannote-Whisper-v3 system. Fine-tuning DiCoW further improves overall performance. Overall, these results demonstrate consistent gains using our diarization-conditioned ASR approach, especially after fine-tuning.

\subsection{Labeling Inconsistencies and Their Impact}\label{sec:annot}
We identify labeling inconsistencies in the training and development splits of the MLC-SLM dataset that affect the performance of both our diarization system, the fine-tuned Pyannote baseline, and likely other participants' systems. To mitigate this, we propose a simple approach using the Silero VAD~\cite{Silero_VAD} model.

The core issue arises from the annotation protocol used in the training and development data. In some cases, actual speech is not annotated and is therefore treated as silence, while in other cases, segments labeled as speech include long internal pauses or silences. This is not problematic in Task~1, where utterance-level ground-truth (GT) segmentation is provided, as each labeled chunk can be processed independently. 

However, in Task~2, which requires joint diarization and ASR, these inconsistencies introduce a mismatch: the diarization system may learn from the MLC-SLM training data to skip valid speech in order to minimize loss, leading to degraded performance on the test set. Unlike the training and development data, the test set follows a cleaner protocol, where all unannotated segments are explicitly muted by the organizers. This, however, introduces a signal-level mismatch relative to the training and development partitions, which can further degrade performance.

To better simulate these test conditions, we create a modified version of the development set in which unannotated segments with noticeable energy are muted, while originally silent regions remain unchanged. Note, however, that the training data used for fine-tuning remains unchanged and inconsistent with the test-like development data.

\begingroup
\setlength{\tabcolsep}{3pt}
\begin{table}[t]
\centering
\caption{Comparison of DER and micro average tcpWER/CER when utilizing out-of-domain (OOD), fine-tuned (FT), and fine-tuned with VAD (FT+VAD) diarization models. Results are shown on the original development data and on a modified “test-like” development set.}
\label{tab:der_wer}
\begin{tabular}{l|ccccc}
\toprule
Diarization Model & DER & Miss & FA & Conf. & tcpWER/CER \\
\midrule
\multicolumn{6}{c}{Original development data}  \\
\midrule
OOD  & 21.8 & 7.5 & 11.5 & 2.8 & 26.7 \\
FT  & 12.7 & 3.8 & 6.7 & 2.2 & 20.8 \\
FT+VAD & 22.8 & 13.4 & 6.9 & 2.5 & 23.6 \\
\midrule
\multicolumn{6}{c}{Test-like development data} \\
\midrule
OOD & 13.5  & 7.5 & 3.1 & 2.9 &19.9 \\
FT & 12.4 & 3.9 & 4.5 & 4.0 & 22.4  \\
FT+VAD & 16.2 & 13.7 & 0.3 & 2.3 & $\mathbf{17.9}$ \\
\bottomrule
\end{tabular}
\vspace{-4mm}
\end{table}
\endgroup
Table~\ref{tab:der_wer} presents diarization performance in terms of DER and tcpWER for both OOD and FT systems on the original development data and a modified ``test-like" version. 

We make two key observations. First, on the original development set, a large portion of the OOD model’s DER comes from false alarms (FA), which often correspond to real speech that is simply unlabeled. This is evident from the FA dropping from 11.5\,\% to 3.1\,\% on the test-like version.

Second, the higher Miss rate of the OOD model on test-like data (7.5\,\% vs. 3.9\,\% for FT) is also misleading. Due to inconsistencies in the original annotations—where some silence regions are labeled as speech—more accurate silence detection appears as an increased Miss rate.

Moreover, the fine-tuned model struggles under test-like conditions, as it often predicts speech in silent regions, causing speaker embeddings to be extracted from silence. This negatively impacts clustering quality and increases speaker confusion errors.
To mitigate these issues, we incorporate a Silero VAD model with a weight of 0.8 to improve speech/silence estimation alongside our diarization system. Additionally, to comply with the annotation protocol—which disallows overlapping speech—we redistribute the probability mass of overlapping frames across individual speakers and select the most likely speaker per frame.
This approach significantly improves tcpWER on the test-like development set from 22.4\,\% to 17.9\,\%, and on the test set from 28.6\,\% to 17.4\,\%.

\begin{figure}[t]
    \centering
    \includegraphics[width=0.95\columnwidth]{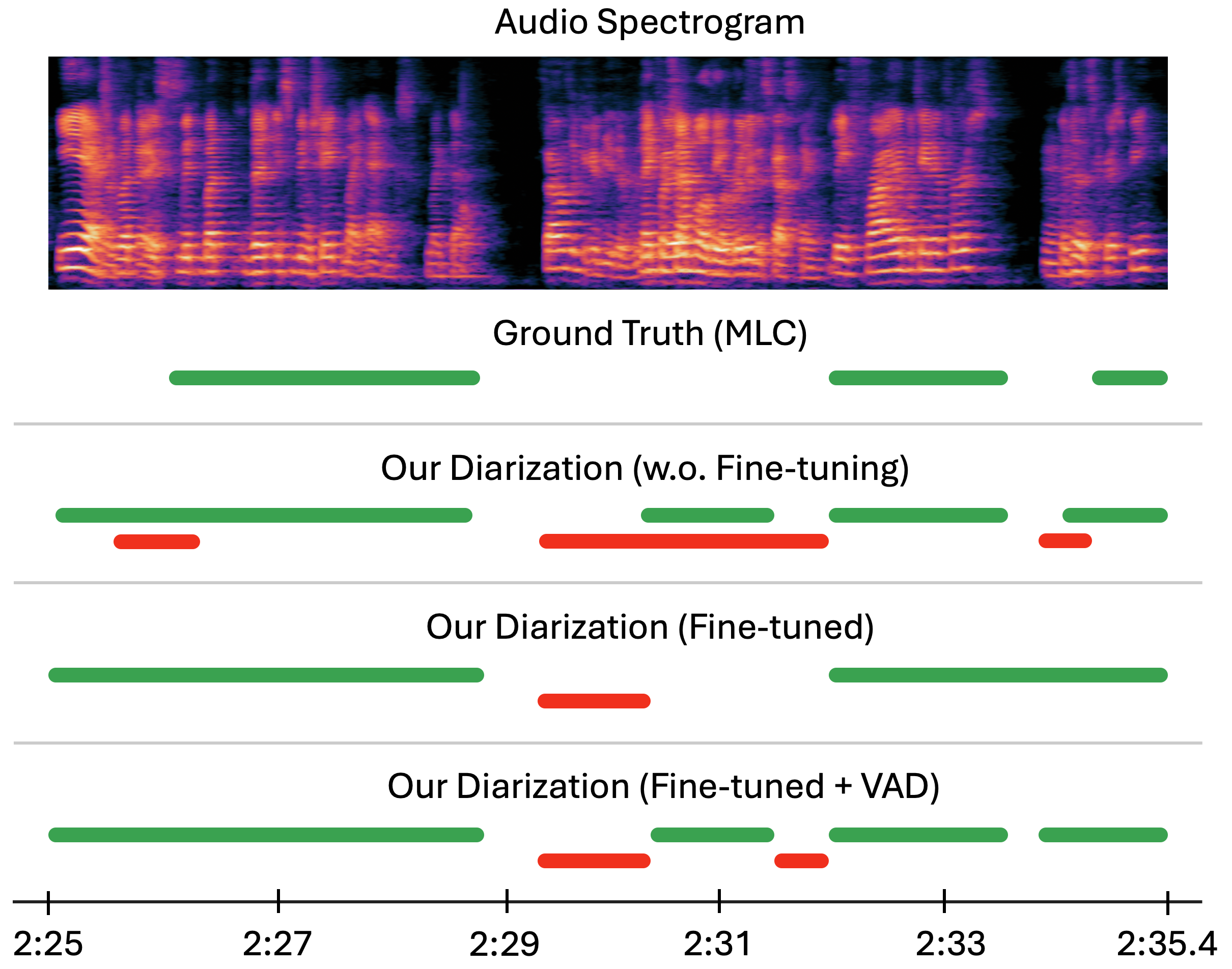}
    \caption{
Example of the ground truth diarization; our system before fine-tuning on MLC; the same system after fine-tuning; and the fine-tuned system with probabilities merged using auxiliary VAD.
    }
\label{fig:example}
\vspace{-5mm}
\end{figure}

Figure~\ref{fig:example} shows an example from the development set (English-American-0525\_002). The difference between the spectrogram and the GT segmentation reveals skipped speech and inaccurate overlaps. On the other hand, the OOD system closely follows actual speech activity, which was confirmed by listening to the audio, while fine-tuning shifts predictions toward inconsistent GT labels. Although FT+VAD seemingly worsens DER, it effectively corrects VAD errors and provides the best performance when used with the TS-ASR system.

\vspace{-2mm}
\section{Conclusions}
We presented a simple, non-LLM approach for multilingual long-form transcription combining diarization and target-speaker ASR. Our system placed second in the MLC-SLM Challenge, showing strong performance across diverse languages.
While effective, our method has limitations that open avenues for future work. In particular, we did not fully explore optimal training strategies for diarization on loosely annotated data. Additionally, replacing DiCoW with a diarization-conditioned speech LLM could further enhance performance and remains a promising direction.

\vspace{-2mm}
\section{Acknowledgements}
The work was supported by Ministry of Education, Youth and Sports of the Czech Republic (MoE) through the OP JAK project ``Linguistics, Artificial Intelligence and Language and Speech Technologies: from Research to Applications" (ID:CZ.02.01.01/00/23\_020/0008518) and by Czech Ministry of Interior project No. VK01020132 ``112". Computing on IT4I supercomputer was supported by MoE through the e-INFRA CZ (ID:90254).

\bibliographystyle{IEEEtran}
\bibliography{mybib}

\end{document}